\newcounter{myctr}
\def\myitem{\refstepcounter{myctr}\bibfont\noindent\ifnum\themyctr>9\else\phantom{0}\fi\hangindent17pt\themyctr.\enskip}

\documentclass{ws-ijqi}
\begin{document}

\markboth{I.~N.~Agafonov, M.~V.~Chekhova, T.~Sh.~Iskhakov,
A.~N.~Penin, G.~O.~Rytikov, and O.~A.~Shumilkina} {Comparative test
of two methods of quantum efficiency absolute measurement}

\catchline{}{}{}{}{}

\title{COMPARATIVE TEST OF TWO METHODS OF QUANTUM EFFICIENCY ABSOLUTE MEASUREMENT BASED ON SQUEEZED VACUUM DIRECT DETECTION\\}

\author{I.~N.~Agafonov, M.~V.~Chekhova, A.~N.~Penin, G.~O.~Rytikov, and O.~A.~Shumilkina}

\address{Physics Department, Moscow State University, Leninskiye Gory\\
Moscow, 119991,
Russia}

\author{T. Sh. Iskhakov}

\address{Max-Planck Institute for the Science of Light, Guenther-Scharowsky-Str. 1 / Bau 24\\
Erlangen,  D-91058, Germany \\
Timur.Iskhakov@mpl.mpg.de}

\maketitle

\begin{history}
\received{Day Month Year}
\revised{Day Month Year}
\end{history}

\begin{abstract}
We realize and test in experiment a method recently proposed for
measuring absolute quantum efficiency of analog photodetectors.
Similarly to the traditional (Klyshko) method of absolute
calibration, the new one is based on the direct detection of
two-mode squeezed vacuum at the output of a traveling wave OPA.
However, in the new method one measures the difference-photocurrent
variance rather than the correlation function of photocurrents
(number of coincidences), which makes the technique applicable for
high-gain OPA.  In this work we test the new method versus the
traditional one for the case of photon-counting detectors where both
techniques are valid.
\end{abstract}


\section{Introduction}
In almost all quantum optic experiments, photodetectors of two types
are used: counting detectors, mainly avalanche photodiodes\cite{A.V.
Burlakov} or photomultipliers, capable to detect single photons, and
analog ones, basically pin photodiodes working with or without
charge integration\cite{Hansen,Yamamoto}. These detectors do not
resolve single photons but can be used for registering intensity
fluctuations. One of the main characteristics of a photodetector is
its quantum efficiency (QE). Since the operation principles of
counting and analog detectors are different, there are two different
definitions of QE. For a counting detector, quantum efficiency is
the ratio of the photocount number to the number of photons incident
on the detector during a certain period of time. For analog
detectors, "the photocount number" in the previous definition should
be replaced by "the number of registered photoelectrons". The
observed QEs for avalanche photodiodes take values up to
76\%\cite{P.G. Kwiat} while the record QE for counting detectors is
88\%\cite{Shigeki Takeuchi}. For the best pin diodes, QE is close to
100\%. Typically, QE measurements are based on reference sources
(black body)\cite{Ott,Ingerson} or reference detectors (blackbody
radiometers)\cite{J.M. Kendall}. In the first case the spectral
radiation density is determined from the known temperature of the
source, in the second one the intensity is determined by the
calorimetric measurement of the heat produced by the beam incident
on the blackened detector of known absorptivity\cite{D.B. Betts}.
\subsection{Klyshko method of QE absolute measurement}
 There is an alternative method of measuring quantum efficiency, which is absolute in the sense
 that it needs no reference sources or detectors. This method, based on the perfect pairwise correlation
 of photons produced by spontaneous parametric down-conversion
 (SPDC),\cite{Zel'dovich,David,Klyshko77,Klyshko80,Malygin81,Klyshko87,Migdall,Ginzburg,Novero,twinphoton06}
 is usually referred to as the Klyshko method of absolute calibration. The key point of the method
 is measuring the rate of photocount coincidences for two detectors that register signal and idler
 radiation at the output of an unseeded nondegenerate optical parametric amplifier (NOPA). The detection of a
 photon in the idler beam guarantees the presence of the photon in the signal one and vice versa. Thus the absence
 of photodetection in one of the channels and its presence in the conjugate one means that the detector has not registered
 photon because of its imperfect quantum efficiency. Since numbers of photons in the signal (s) and idler (i) beams
 are the same and equal to the number of created pairs {\it N},
\begin{equation}
\label{this}
N_i=N_s=N
\end{equation}
there will be $\eta_s N$ and $\eta_i N$ photons detected in the
signal and idler channels, respectively, with $\eta_{s,i}$ denoting
the quantum efficiencies in the signal and idler channels. At the
same time, the number of coincidences is
\begin{equation}
N_c=\eta_i \eta_s N
\end{equation}
The quantum efficiency of one (tested) detector can be determined as the ratio of the coincidence number
and the photocount number in the other (reference) detector:
\begin{equation}
\eta_{s,i}=N_c/N_{i,s}
\end{equation}
In experiment the frequency and angular spectrum registered by the reference detector should be obviously
covered by all conjugate modes collected by the detector under test (DUT). The numbers of accidental coincidences
$N'_c$  and unwanted noises $N'_{i,s}$  (fluorescence, dark noise, etc.) should be also taken into account
as\cite{P.G. Kwiat}
\begin{equation}
\eta_{s,i}=\frac{N_c-N'_c}{N_{i,s}-N'_{i,s}}
\label{QEcor}
\end{equation}
The accuracy of such a measurement reduces when the number of
accidental coincidences is high. Therefore, the method works well
only at low parametric gain, when the normalized second-order
Glauber's correlation function is much larger than unity and,
correspondingly, the number of accidental coincidences is much
smaller than the number of the real ones. In
Ref.~\refcite{S.Polyakov}, it was tested by means of comparison with
another method, using a national primary standard detector scale.
 \subsection{QE absolute measurement based on registering difference-signal variance}
Recently a new universal\footnote{This method works at any
parametric gain and can be used for the calibration of both counting
and analog detectors.} calibration method of photodetectors based on
the registration of two-mode squeezing at the output of a
traveling-wave
OPA~\cite{Jedrkiewicz2004,M.Bondani,IskhakovPRL,Blanchet2008,Brida2009}
was suggested~\cite{Brida2006}. The method is based on the fact that
signal and idler beams emitted via PDC are twins in the sense that
they have identical photon number fluctuations at any parametric
gain. In this context, strong correlations are manifested in the
noise reduction of the intensity difference for signal and idler
beams below the shot-noise level. Noise Reduction Factor ($NRF$) is
used as a quantitative characteristic of two-mode squeezing, which
can be measured in the experiment. This value is defined as
\begin{equation}
\hbox{$NRF$}=\frac{\hbox{Var}(i_{-})}{\langle i_{+}\rangle},
\label{NRFi}
\end{equation}
where $i_{-}=i_s-i_i$ is the difference between photocurrents in the
signal and idler detectors and ${\langle i_{+}\rangle}={\langle
i_i\rangle}+{\langle i_{s}\rangle}$  is the sum of the averaged
signals (usually giving the Shot Noise Level $SNL$ ). In the case of
ideal alignment, $NRF$  does not depend on the parametric gain. It
is important that $NRF$ can be also measured by single-photon
detectors\cite{IskhakovJL}, in which case $i_{-}$ and $i_{+}$ should
be replaced by $N_{-}$ and $N_{+}$,  the difference and sum of the
photocount numbers in the signal and idler detectors, and
Eq.~(\ref{NRFi}) should be modified to
\begin{equation}
NRF=\frac{\hbox{Var}(N_{-})}{\langle N_{+}\rangle}.
\label{NRFN}
\end{equation}
Similarly to the Klyshko absolute calibration method, the
difference-signal method provides the total QE of the optical
channel, $\eta\equiv T\eta_{det}$, where $\eta_{det}$ is the QE of
the detector itself and $T$  is the optical transmission. Under the
assumption that the QEs of the signal and idler optical channels are
the same, $\eta_i=\eta_s\equiv\eta$, $NRF$ is only defined by
$\eta$,
\begin{equation}
NRF=1-\eta.
\label{NRFeta}
\end{equation}
In the case of different quantum efficiencies $\eta_i\neq\eta_s$,
calculation shows that Eq.~(\ref{NRFeta}) becomes more complicated,
\begin{equation}
NRF=1-2\frac{\eta_i\eta_s}{\eta_i+\eta_s}+N\frac{(\eta_i-\eta_s)^2}{\eta_i+\eta_s},
\label{NRFetadif}
\end{equation}
where $N$ is the mean number of photons per mode. In the case of
small-gain parametric down-conversion, the third term in
(\ref{NRFetadif}) can be neglected. In the high-gain regime, the
difference of losses in the two channels leads to $NRF$  dependence
on $N$  and to the reduction of the amount of two-mode
squeezing\cite{IvanN}. This unbalance of the quantum efficiencies
can be canceled by inserting additional losses into one of the
channels (\ref{NRFeta}) or by numerical multiplication of the signal
in the idler channel by a factor $k\equiv\eta_i/\eta_s$,
\begin{equation}
\frac{\hbox{Var}(N_i-kN_s)}{\langle N_i+kN_s\rangle}=
\frac{1+k}{2}-\eta_i,
\end{equation}

The other condition to be satisfied in the detection of twin-beam
squeezing is conjugate multi-mode registration. For PDC radiation,
even an infinitely narrow part of the frequency-angular spectrum of
the signal beam is correlated with a finite frequency-angular
spectrum in the idler beam and vice
versa\cite{Brambilla2004,YuraMV}. Any restriction of the registered
spectrum leads to the existence of unmatched modes which deteriorate
the observable squeezing. Therefore, in order to reduce the
contribution of unmatched modes in comparison with the number of
conjugate ones the detection volumes should be made as large as
possible\cite{YuraMV,IvanN,lastwithall}.

This method of the QE measurement has been first applied to the
calibration of avalanche photodiode in Ref.~\refcite{IskhakovJL}. Recently it was used for the
absolute calibration of a CCD camera\cite{Brida10}.

It is interesting to compare the difference-signal method with the
coincidence (Klyshko) one because totally different conditions
should be satisfied to realize them. For the coincidence method, the
number of collected modes and the parametric gain should be small to
reduce the number of accidental coincidences. On the contrary, for
the difference-signal method, the number of collected modes should
be as large as possible and it can be equally applied at any
parametric gain. The last feature makes this method universal and
allows it to be used for the calibration of both counting and
analogue detectors. Therefore, our comparative test of the two
methods was performed with counting (single-photon) detectors.
\section{Photon-counting detectors and the dead-time effect}
It is worth stressing, however, that special measures were to be
taken to adapt the difference-signal method to the use with
single-photon detectors. An important feature of the latter is the
dead-time (DT) effect\cite{Mandel1979,Sabol}. Namely, after each
photon registration a single-photon detector is unable to register
another photon during a certain period of time, called the dead
time. This effect restricts the maximum counting rate and therefore
influences the photocount statistics. For example, a Poissonian
statistics of photons at the input of a counting detector converts
to a sub-Poissonian statistics of photocounts at the output. So if
there is an n-photon pulse, with the duration smaller than the DT,
at the input of a single-photon detector with quantum efficiency
$\eta$, there will be only one ``click'' at the output, with the
probability $a_n=1-(1-\eta)^n$. For PDC, the probability
distribution of photon numbers in signal and idler beams is thermal,
but if a large number of modes is registered, photon-number
distribution for the whole ensemble is Poissonian,
$P(n)\equiv\frac{\langle N_p\rangle^ne^{-\langle N_p\rangle}}{n!}$,
with the mean photon number $\langle N_p\rangle$. The probability of
a photocount is then
\begin{equation}
p=\sum\limits_{n}a_nP(n).
\end{equation}
Taking into account this formula, as well as the fact that the
intensities in the signal and idler beams are absolutely correlated,
we find the value of $NRF_{meas}$ according to (\ref{NRFN}). Up to
the terms linear in the mean photon numbers per pulse in the signal
and idler detectors, $\langle N_{s,i}\rangle$, the measured value of
$NRF$ is
\begin{equation}
NRF_{meas}=1-\frac{2\eta_i}{1+k}-\langle
N_+\rangle\left[\frac{1+k^2}{(1+k)^2}-\eta_i\frac{k^2+4k+1}{(1+k)^3}+\eta_i^2\frac{1}{(1+k)^2}\right],
\label{NRFlarge}
\end{equation}
where $\langle N_{+}\rangle\equiv\langle N_{i}\rangle+\langle
N_{s}\rangle$.

We see that the DT effect can be only negligible at $\langle
N_+\rangle\ll 1$. However, working at extremely low signal levels,
such as $\langle N_+\rangle\approx 10^{-3}$ or less, is technically
difficult as it requires large acquisition times. In our experiment,
we worked at $\langle N_+\rangle\approx 5\cdot10^{-2}$, and the DT
effect had to be taken into account. This was performed by measuring
$NRF$  and $k$ and then solving Eq.~(\ref{NRFlarge}) for $\eta_i$
(only the positive root of the quadratic equation was taken into
account).

In the coincidence method, the DT effect leads to the
underestimation of both coincidences and singles numbers in
Eq.~(\ref{QEcor}). As the reference detector selects much narrower
frequency and angular bands than the DUT, its mean number of counts
per pulse is usually small (in our case, it was on the order of
$10^{-3}$), and the denominator in Eq.~(\ref{QEcor}) did not require
any DT corrections. At the same time, the DUT mean number of counts
per pulse was typically as high as $3\cdot10^{-2}$, and its DT
effect led to the reduction of the coincidence counting rate. The
same simple model as we use to describe the DT effect on the $NRF$
predicts the reduction of measured mean coincidence number per pulse
by a value of $\langle N_{c}\rangle\langle N_{+}\rangle$, where
$\langle N_{c}\rangle$ is the mean number of real coincidences per
pulse. In all measurements of QE using Eq.~(\ref{QEcor}), this
correction was taken into account.
\section{Experiment}
In the comparative test of the difference-signal method and the
coincidence-counting one, we used the experimental setup shown in
Fig.~1. PDC was generated in a 3 mm $LiIO_3$ crystal by the third
harmonic of a pulsed Nd:YAG laser with the wavelength 355 nm, pulse
duration 5 ns, and repetition rate 10kHz. The beam waist diameter in
the crystal was $0.4$ mm. The crystal was cut for type-I phase
matching and oriented to generate signal and idler radiation at
wavelengths 650 nm and 780 nm in the collinear direction. The
parametric gain could be varied by rotating the half-wave plate in
front of a Glan prism (GP) in the pump beam. In our experiment, it
took values of up to $10^{-4}$, so that the spontaneous regime of
PDC was realized. To eliminate the fundamental and second-harmonic
radiation, as well as the fluorescence of the optical elements in
the pump beam, a dispersive prism in combination with an aperture
and an ultraviolet filter (UVG1) were used. Another ultraviolet
filter (UVG2) was inserted to eliminate the Glan prism fluorescence.
After the crystal, a dichroic mirror UVM and a white-glass filter
(WG) cut off the radiation of the pump and transmitted the SPDC
radiation. The fluorescence of the crystal was reduced by the
red-glass filter RG. Signal and idler beams were separated by a
dichroic beamsplitter (DBS) and focused by lenses on the detectors,
which were avalanche photodiodes (``Perkin\&Elmer'' APD Ñ30902S)
operating in the Geiger mode with passive quenching. The angular
spectrum detected in each channel was restricted by iris apertures
(A1, A2) inserted at the outputs of the DBS. All optical elements
after the crystal had antireflection coating within the band $600$
nm - $800$ nm to reduce the losses. The signals from the detectors
were processed in the registration part of the setup consisting of
two counters (N1, N2) and the coincidence circuit (CC) with the time
resolution 4.2 ns. The counters were gated by pulses of duration 30
ns, synchronized with the pump pulses. As a result, the dark count
noise was reduced to the level of $10^{-5}$ photons per gate. At the
output, the registration system provided the numbers of photocounts
from the detectors as well as the numbers of coincidences during
each laser pulse. The average photocount number per pulse did not
exceed 0.03 but even in this case the DT effect had a certain
influence on the method.


\begin{figure}
\begin{center}
\includegraphics[width=0.75
\textwidth]{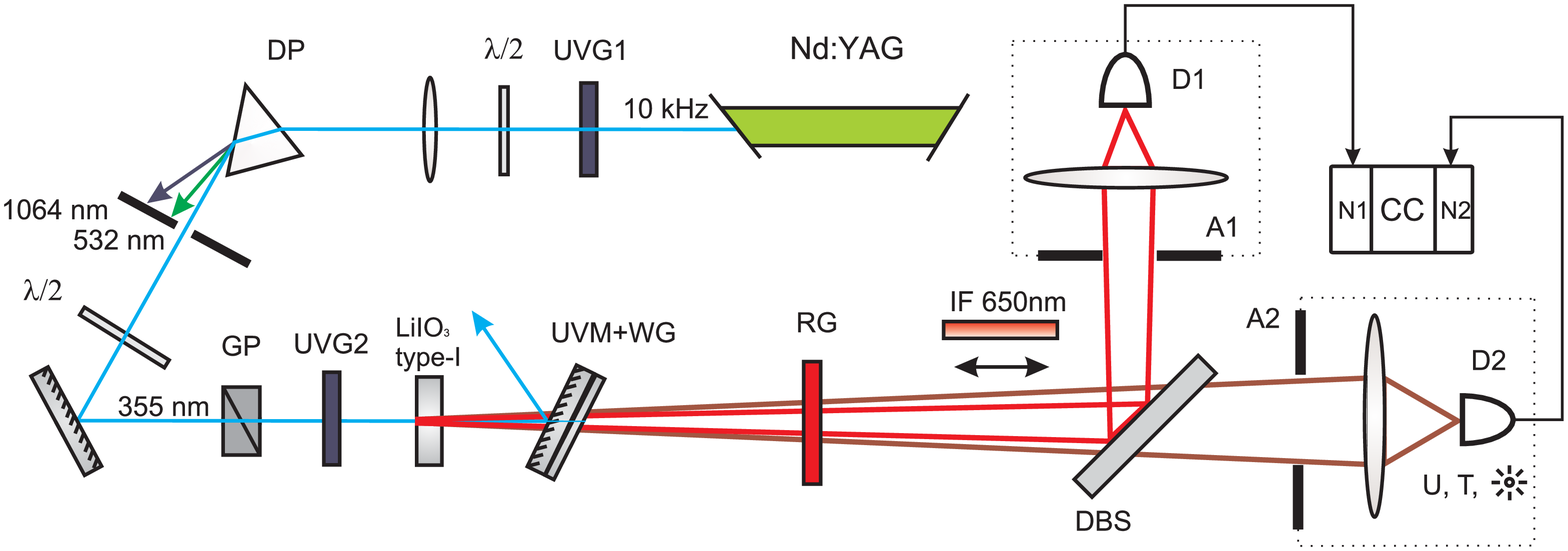} \caption{Experimental setup. UVG1, UVG2,
ultraviolet filters; DP, dispersive prism; GP, Glan prism; UVM,
ultraviolet mirror; WG, white glass filter; RG, red glass filter;
DBS, dichroic beamsplitter; A1 and A2, apertures; IF 650 nm,
interference filter with bandwidth 10 nm centered at 650 nm; D1 and
D2, counting detectors;  N1 and N2, counters working in the gated
mode; CC, coincidence circuit.}\end{center} \end{figure}

Using simple modifications of this setup it was possible to realize
both calibration methods for the detector registering idler
radiation (at the wavelength $780$ nm). To measure QE by the
coincidence counting method, in the reference (signal) channel an
interference filter was inserted with the bandwidth 10 nm centered
at 650 nm (IF 650nm) and the diameter of the A2 aperture was 2 mm.
In the channel under test, the aperture size was 8 mm and no filter
was inserted. The distance from the crystal to the apertures was 108
cm. The quantum efficiency was determined from Eq.~(\ref{QEcor}),
with a correction for the dead-time effect. The number of accidental
coincidences was found as\cite{O.A. Ivanova} $N'_c=N_1N_2K$, where
$K=0.65$. To measure the noise in the reference channel, $N'$, the
orientation of the crystal was changed to eliminate the SPDC.

For the difference-signal method, the aperture diameters in the
signal and idler channels were matched to satisfy the transverse
phasematching condition for PDC\cite{IvanN}:
\begin{equation}
\frac{D_i}{D_s}=\frac{\lambda_i}{\lambda_s}.
\label{pm}
\end{equation}
As a result, conjugated modes of the signal and idler radiation were
registered. In the experiment, the aperture diameters were chosen to
be 5 and 6 mm. The QE was calculated with an account for the DT
effect, according to Eq.~(\ref{NRFlarge}).

In the first measurement, we checked the validity of
Eq.~(\ref{NRFlarge}) by measuring $NRF$ versus the mean number of
photocounts per pulse, which was changed by varying the parametric
gain. Dependence of $NRF$ on the sum signal in the two channels is
presented in Fig.~2. Since the variance of the difference photon
number for SPDC, according to the theory, does not depend on the
gain, the observed $NRF$ decrease with the growth of the signal is
only due to the DT effect. Solid line is a fit plotted according to
(\ref{NRFlarge}) with the only fitting parameter $\eta_i$. The
parameter $k\equiv\eta_i/\eta_s$ was measured independently, from
the ratio of the signals in the channels.
\begin{figure}
\begin{center}
\includegraphics[width=0.75
\textwidth]{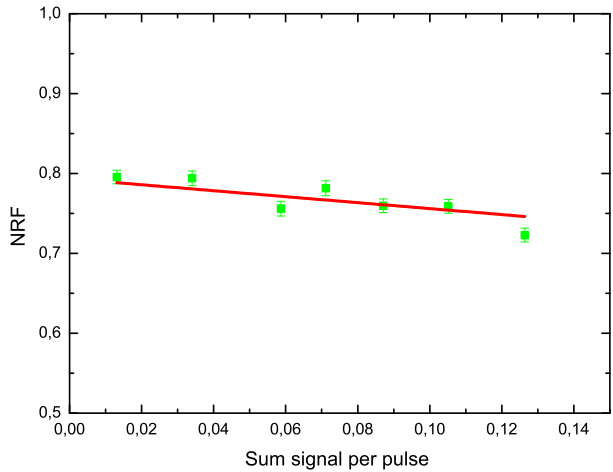} \caption{ Influence of the dead-time effect on
the difference-signal variance measurement. $NRF$ is plotted as a
function of the sum signal of the two detectors varied through
changing the pump intensity.}\end{center}
\end{figure}

For the comparison of two calibration methods, QE was measured as a
function of the main parameters of the detectors, such as the
temperature and the wavelength. In addition, some `external'
parameters were added: losses in the optical channel and a strong
background noise. All measurements were carried out with the other
parameters kept constant. Unless stated otherwise, the temperature
of the detector was $-25^{\circ}C$ and the bias voltage was $15$V
above the breakdown one. With these parameters fixed, the quantum
efficiency was measured by both methods. The resulting values were
$\eta_0^{c}=0.258\pm 0.004$ from the coincidence method and
$\eta_0^{d}=0.256\pm 0.004$ from the difference-signal method, which
is in perfect agreement. The averaged value
$\eta_0\equiv(\eta_0^{c}+\eta_0^{d})/2$, was used further for
plotting the theoretical dependencies.

Dependence of the QE on the transmission of the optical channel $T$
obtained by the difference-signal method (squares) and the
coincidence counting one (circles) is shown in Fig.~3. Variable
losses were produced by a polarization filter inserted in front of
the DUT. Transmission of the optical channel was estimated using the
output DUT signal. Within the limits of accuracy, the measured
points are in good agreement with each other. They are also in
agreement with the theoretical dependence, $\eta=\eta_0 T$, shown in
the figure by a solid line.

\begin{figure}
\begin{center}
\includegraphics[width=0.75
\textwidth]{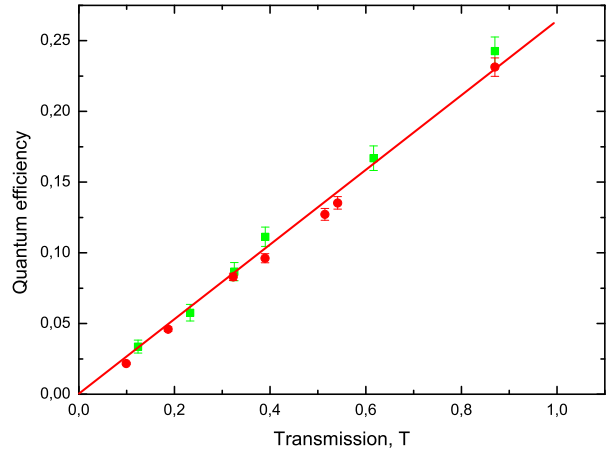} \caption{Quantum efficiency measured
by the difference-signal method (squares) and the coincidence
counting one (circles) as a function of the channel transmission
$T$. Solid line: theoretical dependence.}\end{center} \end{figure}

Dependence of the QE on the temperature is shown in Fig.~4. For each
point, the bias voltage was changed to keep the same level $(15V)$
over the breakdown voltage, as the latter increases with the
temperature. Under these conditions, the datasheets predict no
dependence of the QE on the temperature\cite{datasheet}.  In good
agreement with this, both methods show no temperature dependence
(Fig.~4), and the data are in a satisfactory agreement with each
other. Solid line shows the expected value of the QE, $\eta=\eta_0$,
for all temperatures.
\begin{figure}
\begin{center}
\includegraphics[width=0.75
\textwidth]{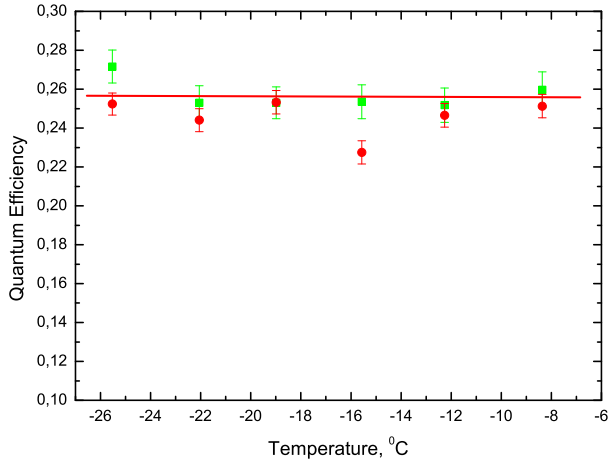} \caption{Quantum efficiency measured by
the difference-signal method (squares) and the coincidence counting
method (circles) as a function of the temperature.}\end{center}
\end{figure}

\begin{figure}
\begin{center}
\includegraphics[width=0.75
\textwidth]{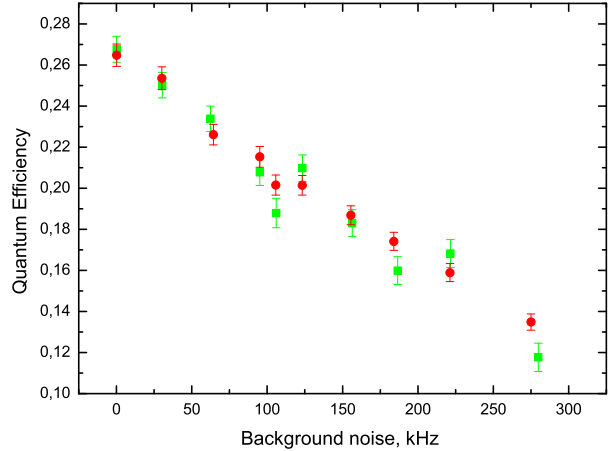} \caption{Quantum efficiency measured by the
difference-signal method (squares) and the coincidence counting one
(circles) as a function of the background noise.}\end{center}
\end{figure}
Figure 5 shows the influence of the external continuous noise on the
QE measured by two different methods. The external noise reduces the
QE due to the DT effect, which leads to the saturation of the
detector; the QE measurement is possible since the noise is
continuous while the signal is pulsed. A daylight bulb was used as a
continuous external noise source. In addition to reducing the QE,
the external noise added to the signal measured during $30$-ns
gates, but this additional noise was taken into account by making a
separate measurement, with the signal eliminated. The external noise
radiation was found to be weak enough to be shot-noise limited.
Taking  into account all this, as well as the fact that the noise
and the SPDC radiation are statistically independent, we can write
Eq. (\ref{NRFN}) as
\begin{equation}
\frac{\hbox{Var}(N_{-})}{\langle
N_{+}\rangle}=\frac{\hbox{Var}(N_{-})_{S}+\hbox{Var}(N_{-})_{N}}{\langle
N_{+}\rangle_{S}+\langle N_{+}\rangle_{N}}, \label{NRFnoise}
\end{equation}
where $\hbox{Var}(N_{-})_{S}$, $\langle N_{+}\rangle_{S}$ and
$\hbox{Var}(N_{-})_{N}$, $\langle N_{+}\rangle_{N}$ are the
variances of the difference signal and the mean sum signals for PDC
and the external noise, respectively. For measuring the QE by the
difference-signal method, the values of $\hbox{Var}(N_{-})_{N}$,
$\langle N_{+}\rangle_{N}$ were subtracted  from the numerator and
denominator in the left-hand side of (\ref{NRFnoise}). In the
coincidence method, the QE value was calculated from (\ref{QEcor})
taking into account the number of accidental coincidences and the
background noise. This dependence is of practical use as it
demonstrates the QE reduction and the consequent reduction of the
communication speed in systems based on counting detectors. Fig.5
shows a good agreement between the data obtained by both methods.

Finally, we have applied the two methods to the measurement of QE
spectral dependence. In this case, every point requires a different
alignment of the setup. Indeed, the orientation of the crystal
should be adjusted each time to provide the phase matching at the
corresponding wavelengths. In addition, the difference-signal method
implies condition (\ref{pm}), and hence the iris diameters should be
chosen different for every point of the spectral dependence. The
results of the QE measurement in the range from $780$ nm to $860$ nm
are shown in Fig.~6. For aligning the crystal, a narrowband
interference filter was inserted into the reference channel, and
then the QE was measured by means of the coincidence method. After
that, the filter was removed, the diameters of the irises were set
according to Eq.~(\ref{pm}), and the QE was measured by means of the
difference-signal method. At the wavelength $866$ nm, only the
coincidence counting method was applied as the dichroic beamsplitter
did not well separate the signal and idler beams (transmitted too
much of the signal radiation). At $850$ nm, the QE was measured only
by the difference-signal method because of the absence of the
corresponding narrowband filter at the signal wavelength. The
measurements by two different methods are in a satisfactory
agreement with each other and the data presented in the
datasheet\cite{datasheet}.
\begin{figure}
\begin{center}
\includegraphics[width=0.75
\textwidth]{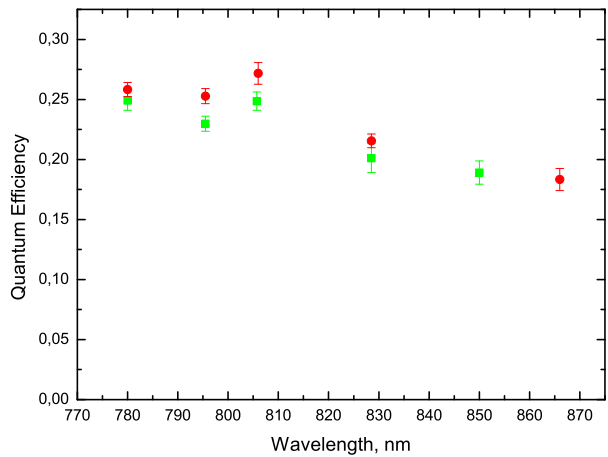} \caption{Quantum efficiency measured by
the difference-signal method (squares) and the coincidence counting
method (circles) as a function of the wavelength.}\end{center}
\end{figure}

In conclusion, we have tested the recently proposed method of
absolute quantum efficiency measurement based on twin-beam
squeezing. Our test relied on the comparison between this method and
the well-known absolute calibration method based on registering
coincidences of photocounts. Since the latter works well only for
photon-counting detectors, both methods were realized in a setup
with low-gain parametric down-conversion and single-photon
detectors. However, measurement of twin-beam squeezing has certain
difficulties in the case of single-photon detectors, and one of the
results of our work is developing a technique for measuring the
difference-signal variance with single-photon detectors. In
particular, the dead-time effect in single-photon detectors was
taken into account. The obtained results demonstrate both the
validity of the method based on the difference-signal variance and
its applicability to single-photon detectors. Dependencies of
quantum efficiency on the additional losses, wavelength,
temperature, and external noise (which influences quantum efficiency
due to the dead-time effect) were obtained.

\section*{Acknowledgments}

This work was supported by the Russian Foundation for Basic
Research, grants \#\# 08-02-00555, 08-02-00741, 10-02-00202.
Theoretical help by A.~L.~Chekhov in taking into account the
dead-time effect is gratefully acknowledged.


\begin{thebibliography}{20}


\bibitem{A.V. Burlakov} A.~V.~Burlakov, C.~Novero, A passively quenched avalanche
photodiode for single photon detection, {\it R.T.630} (2001).

\bibitem{Hansen} H.~Hansen, T.~Aichele, C.~Hettich, P.~Lodahl, A.~I.~Lvovsky, J.~Mlynek, S.~Schiller,
An ultra-sensitive pulsed balanced homodyne detector: application to time-domain quantum measurements,
{\it Opt., Lett.} {\bf 26}  (2001) 1714--1716.

\bibitem{Yamamoto} Y.~Yamamoto, N.~Imoto, S.~Machida, Amplitude squeezing in a semiconductor laser
using quantum nondemolition measurement and negative feedback,{\it Phys. Rev. A} {\bf 33} (1986) 3243.

\bibitem{P.G. Kwiat} P.~G.~Kwiat, A.~M.~Steinberg, R.~Y.~Chiao, P.~H.~Eberhard, M.~D.~Petroff, High-efficiency
single-photon detectors, {\it Phys. Rev. A} {\bf 48} (1993) R867--R870.

\bibitem{Shigeki Takeuchi} S.~Takeuchi, J.~Kim ,Y.~Yamamoto, H.~H.~Hogue, Development of a high-quantum-effciency
single-photon counting system, {\it Appl. Phys. Lett.} {\bf 74} (1999) 1063.

\bibitem{Ott} W.~R.~Ott, P.~Fieffe-Prevost, and W.~L.~Wiese,  VUV Radiometry with Hydrogen Arcs.
1: Principle of the Method and Comparisons with Blackbody
Calibrations from 1650{\AA} to 3600{\AA}, {\it Applied Optics} {\bf
12}, Issue 7 (1973) 1618--1629.

\bibitem{Ingerson} T.~E.~Ingerson, R.~J.~Kearney, and R.~L.~Coulter, Photon counting with photodiodes,
{\it Appl. Opt.} {\bf 22} (1983) 2013--2018.

\bibitem{J.M. Kendall} J.~M.~Kendall, Sr. and C.~M.~Berdahl, Two Blackbody Radiometers of High Accuracy,
{\it Appl. Opt.} {\bf 9}, Issue 5 (1970) 1082-1091.

\bibitem{D.B. Betts} D.~B.~Betts, E.~J.~Gillham, An International Comparison of Radiometric Scales,
{\it Metrologia} {\bf 4}, (1968) 101.

\bibitem{Zel'dovich} C.~Y.~Zel'dovich and D.~N.~Klyshko, Statistics of feld in parametric luminescence,
{\it Sov. Phys. JETP Lett.} {\bf 9} (1969) 40--44.

\bibitem{David} D.~C.~Burnham and D.~L.~Weinberg, Observation of Simultaneity in Parametric Production of Optical
Photon Pairs, {\it Phys. Rev. Lett.} {\bf 25} (1970) 84--87.

\bibitem{Klyshko77} D.~N.~Klyshko, {\it Sov. J. Quantum Electron} {\bf 7} (1977) 591.

\bibitem{Klyshko80} D.~N.~Klyshko, Use of two-photon light for absolute calibration of photoelectric detectors,
{\it  Sov. J. Quantum Electron.} {\bf 10} (1980) 1112--1116.

\bibitem{Malygin81} A.~A.~Malygin, A.~N.~Penin, and A.~V.~Sergienko, Absolute calibration of the sensitivity of
photodetectors using a two-photon field, {\it Sov. Phys. JETP Lett.} {\bf 33} (1981) 477--480.

\bibitem{Klyshko87} D.~N.~Klyshko and A.~N.~Penin, The prospects of quantum photometry, {\it Sov. Phys. Usp.}
{\bf 30} (1987) 716--723.

\bibitem{Migdall} A.~Migdall, Correlated-Photon Metrology Without Absolute Standards,
{\it Phys. Today} {\bf 52} (1999) 41--46.

\bibitem{Ginzburg} V.~M.~Ginzburg, N.~G.~Keratishvili, E.~L.~Korzhenevich, G.~V.~Lunev, A.~N.~Penin,
and V.~I.~Sapritsky, Absolute meter of photodetector quantum
efficiency based on the parametric down-conversion effect, {\it Opt.
Eng.} {\bf 32} (1993) 2911--2916.

\bibitem{Novero} G.~Brida, M.~Genovese, and C.~Novero, An application of two photon entangled states to
quantum metrology, {\it J. Mod. Opt.} {\bf 47} (2000) 2099--2104.

\bibitem{twinphoton06} G.~Brida, M.~Genovese, M.~Gramegna,Twin-photon techniques for photo-detector calibration, {\it Laser Physics Letters} {\bf 3} no. 3 (2006) 115--123.

\bibitem{S.Polyakov} S.~V.~Polyakov and A.~L.~Migdall, High accuracy photon-counting detector calibration
and independent verification of a correlated-photon calibration
technique, {\it Optics Express} {\bf 15}, Issue 4 (2007) 1390--1407.

\bibitem{Jedrkiewicz2004} O.~Jedrkiewicz, Y.-K.~Jiang, E.~Brambilla, A.~Gatti, M.~Bache, L.~A.~Lugiato,
P.~Di Trapani, Detection of sub-shot-noise spatial correlations in
high-gain Parametric Down Conversion, {\it Phys. Rev. Lett.} {\bf
93} (2004) 243601.

\bibitem{M.Bondani}  M.~Bondani, A.~Allevi, G.~Zambra, M.~G.~A.~Paris, A.~Andreoni,
Sub-shot-noise photon-number correlation in mesoscopic twin-beam of light, {\it Phys. Rev. A} {\bf 76} (2007) 013833.

\bibitem{IskhakovPRL} T.~Sh.~Iskhakov, M.~V.~Chekhova, G.~Leuchs, Generation and direct detection
of broadband mesoscopic polarisation-squeezed vacuum, {\it Phys. Rev. Lett.} {\bf 102} (2009) 183602.

\bibitem{Blanchet2008} J.-L.~Blanchet, F.~Devaux, L.~Furfaro, and E.~Lantz, Measurement of
Sub-Shot-Noise Correlations of Spatial Fluctuations in the
Photon-Counting Regime, {\it Phys. Rev. Lett.} {\bf 101} (2008)
233604.

\bibitem{Brida2009} G.~Brida, L.~Caspani, A.~Gatti, M.~Genovese, A.~Meda,
  and I.~Ruo-Berchera, Measurement of Sub-Shot-Noise Spatial Correlations without Background Subtraction,
  {\it Phys. Rev. Lett.} {\bf 102} (2009) 213602.

\bibitem{Brida2006} G.~Brida, M.~Genovese, I.~Ruo-Berchera, M.~V.~Chekhova, and A.~N.~Penin, Possibility of
absolute calibration of analog detectors by using parametric
downconversion: a systematic study, {\it JOSA B} {\bf 23}, Issue 10
(2006) 2185--2193; G.~Brida, M.~Chekhova,  M.~Genovese, A.~Penin,
M.L.~Rastello and I.~Ruo-Berchera, Absolute calibration of analog detectors by using parametric
down conversion, {\it IEEE Trans. I.}\&{\it M.} {\bf 56} (2007) 275; G.~Brida, M.~Chekhova, M.~Genovese, I.~Ruo-Berchera, Analysis of the possibility of analog detectors calibration by exploiting stimulated parametric down conversion, {\it Optics Express} {\bf 16} (2008) 12550--12558; G.~Brida, M.~Chekhova, M.~Genovese, M.L.~Rastello, I.~Ruo-Berchera, Absolute calibration of Analog Detectors by using Stimulated Parametric Down Conversion, {\it Journal of Modern Optics} {\bf 56}  (2009) 401--404.

\bibitem{IskhakovJL} T.~Sh.~Iskhakov, E.~D.~Lopaeva, A.~N.~Penin, G.~O.~Rytikov, and M.~V.~Chekhova, Two Methods
for Detecting Nonclassical Correlations in Parametric Scattering of Light, {\it JETP Lett.} {\bf 88} (2008) 660--664.

\bibitem{IvanN} I.~N.~Agafonov, M.~V.~Chekhova, G.~Leuchs, Two-Color Bright Squeezed Vacuum, {\it Phys. Rev.
A}, to appear (2010).

\bibitem{Brambilla2004} E.~Brambilla, A.~Gatti, M.~Bache, and L.~A.~Lugiato,
Simultaneous near-feld and far-feld spatial quantum correlations in
the high-gain regime of parametric down-conversion, {\it Phys. Rev.
A} {\bf 69} (2004) 023802.

\bibitem{YuraMV} G.~O.~Rytikov and M.~V.~Chekhova, Detection of two-mode compression and degree of entanglement
in continuous variables in parametric scattering of light, {\it Jetp} {\bf 134}, No. 6 (2008) 1082--1092.

\bibitem{lastwithall} I.~Agafonov, M.~Chekhova, T.~Iskhakov and G.~Leuchs, Multimode Detection of Broadband Squeezed
Vacuum,{\it Proceedings of NATO Advanced Research Workshop 'Quantum
Cryptography and Computing: Theory and Implementation', Gdansk,
Poland, 9--12 September (2009)}.

\bibitem{Brida10} G.~Brida, I.~Degiovanni, M.~Genovese,  M.~Rastello, I.~Ruo-Berchera, Detection of multimode spatial correlation in PDC and application to the absolute calibration of a CCD camera, {\it arXiv 1005.2937}.

\bibitem{Mandel1979}  L.~Mandel, Inversion problem in photon counting with dead time,
{\it JOSA} {\bf 70}, Issue 7 (1979) 873--874.

\bibitem{Sabol} J.~Sabol, Dead-time Corrections and Effects of Dead-Time on the Counting Statistics of
G-M Counters Used in Mixed Neutron-Gamma Radiation Dosimetry, {\it
Radiation  Protection  Dosimetry} {\bf 23}, Issue 1-4 (1988)
445--447.

\bibitem{O.A. Ivanova}  O.~A.~Ivanova, T.~Sh.~Iskhakov, A.~N.~Penin, M.~V.~Chekhova, Multiphoton
correlations in parametric down-conversion and their measurement in
the pulsed regime,{\it Quantum Electronics} {\bf 36}, Issue 10
(2006) 951--956.

\bibitem{datasheet} Product data sheets C30902E, C30902S, C30921E, C30921S, PerkinElmer
Optoelectronic.

\end{thebibliography}
\end{document}